\documentclass[twocolumn]{aastex631}
\usepackage{kotex}
\usepackage{graphicx}
\usepackage{float}
\usepackage{hyperref}

\newcommand{\HI}{\hbox{{\rm H}\kern 0.2em{\sc i}}}

\begin{document}

\title{Searching for Dark Galaxies with HI detection from the Arecibo Legacy Fast ALFA (ALFALFA) survey}

\correspondingauthor{Ho Seong Hwang}
\email{hhwang@astro.snu.ac.kr}

\author[0009-0007-9451-5337]{Minseong Kwon}
\affiliation{Astronomy Program, Department of Physics and Astronomy, Seoul National University, 1 Gwanak-ro, Gwanak-gu, Seoul 08826, Republic of Korea}

\author[0000-0003-3428-7612]{Ho Seong Hwang}
\affiliation{Astronomy Program, Department of Physics and Astronomy, Seoul National University, 1 Gwanak-ro, Gwanak-gu, Seoul 08826, Republic of Korea}
\affiliation{SNU Astronomy Research Center, Seoul National University, 1 Gwanak-ro, Gwanak-gu, Seoul 08826, Republic of Korea}
\affiliation{Australian Astronomical Optics - Macquarie University, 105 Delhi Road, North Ryde, NSW 2113, Australia}

\author[0000-0002-8990-1811]{Brian R. Kent}
\affiliation{National Radio Astronomy Observatory, 520 Edgemont Road, Charlottesville, VA 22903, USA}

\author[0000-0001-9163-0064]{Ilsang Yoon}
\affiliation{National Radio Astronomy Observatory, 520 Edgemont Road, Charlottesville, VA 22903, USA}

\author[0009-0007-8443-3143]{Gain Lee}
\affiliation{Astronomy Program, Department of Physics and Astronomy, Seoul National University, 1 Gwanak-ro, Gwanak-gu, Seoul 08826, Republic of Korea}

\author{Hyein Yoon}
\affiliation{
Institute for Data Innovation in Science, Seoul National University, 1 Gwanak-ro, Gwanak-gu, Seoul 08826, Republic of Korea}
\affiliation{Astronomy Program, Department of Physics and Astronomy, Seoul National University, 1 Gwanak-ro, Gwanak-gu, Seoul 08826, Republic of Korea}
\affiliation{Sydney Institute for Astronomy, School of Physics A28, The University of Sydney, NSW 2006, Australia}

\begin{abstract}
We present a catalog of 142 dark galaxy candidates in a region covered by the Arecibo Legacy Fast ALFA (ALFALFA) survey. We start with 344 ALFALFA {\HI} sources without optical counterparts and remove those that do not seem to have dark galaxy origin. To do that, we first eliminate 83 sources that are known {\HI} clouds probably formed from tidal interactions between galaxies and 13 sources that have optical counterparts. We then remove 56 sources located near other {\HI} sources, which are likely to be {\HI} clouds. We further exclude 10 sources that have nearby {\HI} sources within the ALFALFA beam and 40 sources potentially associated with nearby galaxies.  We perform visual inspection of optical images from DESI Legacy Imaging Survey with an $r$-band surface brightness limit of $\sim 28.5 \rm \ mag\ arcsec^{-2}$ as well as NUV images from GALEX to confirm the absence of stellar emission. We additionally inspect infrared images from WISE and AKARI for dust emission. As a result, we identify 142 dark galaxy candidates and analyze their physical properties by comparing with luminous galaxies. 
We find that the dark galaxy candidates generally have smaller dynamical masses, higher {\HI}-to-dynamical mass ratios, and are located in less dense regions when compared to luminous galaxies, which is consistent with results from cosmological simulations. This sample provides an important testbed for studying the role of dark matter in galaxy formation and evolution.
\end{abstract}

\keywords{catalogs, {\HI} line, dark matter, dwarf galaxies, galaxy formation, galaxy evolution}

\section{Introduction} \label{sec:intro}
The standard cosmological model, $\rm \Lambda CDM$ (cold dark matter with cosmological constant $\rm \Lambda$\footnote{A recent study based on a tomographic Alcock–Paczyński test with redshift-space correlation function suggests that a model more general than $\rm \Lambda CDM$ (i.e. wCDM model where w is the dark energy equation of state) is required to explain the observed large-scale structures in the universe \citep{2023ApJ...953...98D}}), has successfully explained the physical properties of large scale structures in the universe \citep[e.g.][]{2012ApJ...759L...7P,2016ApJ...818..173H}.
The $\rm \Lambda CDM$ model predicts hierarchical galaxy formation, where small dark matter halos formed from local overdensities gradually merge to create larger halos. Massive halos are expected to grow hierarchically (i.e. host galaxies), and form stars in their deep potential wells. However, during this process, some galaxies may have few or no stars, becoming optically invisible for various reasons; these are referred to as ``dark galaxies'' \citep[e.g.][]{1997MNRAS.292L...5J,2001MNRAS.322..658T,2007ApJ...665L..15K}.

There have been several studies from simulations trying to understand the physical mechanism behind the formation of dark galaxies. \cite{2017MNRAS.465.3913B} investigated low-mass halos using the APOSTLE zoom-in simulations and found that dark galaxies in the local universe (referred to as RELHICs in their terminology) are distinct from most Ultra Compact High Velocity Clouds (UCHVCs). \cite{2020MNRAS.498L..93J} suggested that dark galaxies have higher spins, larger sizes and lower surface densities, which can lead to lower star formation rates.

Recently, \cite{2024ApJ...962..129L} used the IllustrisTNG simulation \citep{2018MNRAS.475..624N,2018MNRAS.477.1206N,2018MNRAS.480.5113M,2018MNRAS.475..676S,2018MNRAS.475..648P} and found that the majority of dark galaxies are located in less dense regions compared to luminous galaxies in the present epoch without any star-forming gas. This result suggests that dark galaxies originate from regions that are different from those of luminous galaxies (i.e. different initial conditions); they are formed in under-dense regions (which prevents star formation due to UV heating from cosmic reionization) and have undergone few interactions or mergers with other systems.

Understanding the physical mechanisms behind the formation of dark galaxies can provide important clues addressing small scale problems of the $\rm \Lambda CDM$ galaxy formation model \citep[e.g.][]{2017ARA&A..55..343B}.
One of these problems is the ``core-cusp" problem, which refers to the difference in the density profile at the center of galaxies between simulations and observations \citep{1994Natur.370..629M, 2015AJ....149..180O}. 
One possible explanation for this problem is baryonic feedback in the form of supernova explosions \citep{2013MNRAS.429.3068T, 2014ApJ...793...46O}. However, implementing realistic baryonic effects in numerical simulations has not been easy \citep[e.g.][]{2023ARA&A..61..473C}. Dark galaxies can be useful tools for studying the effects of baryonic physics. Another problem of the $\Lambda \rm CDM$ model is the mismatch between the number of low-mass satellite galaxies around massive ones in observations and simulations: i.e. missing satellite problem \citep{1999ApJ...522...82K,1999ApJ...524L..19M, 2002MNRAS.336..541V, 2018NatAs...2..162P}.
This discrepancy can also be mitigated by incorporating baryonic physics into numerical simulations \citep[e.g.][]{2016ApJ...827L..23W,2024ApJ...964..123J}; some halos are unable to form stars because the conditions required for star formation are not fulfilled, and some are destroyed by the tidal effects of the stellar disk of the host galaxy. Thus, the number of dark matter halos hosting stars is significantly reduced, and there should be dark matter only halos (i.e. dark galaxies) without stars.

Despite the absence of stars, dark galaxies can be detected by 21-cm neutral hydrogen ({\HI}) emission. Recent studies have shown that neutral hydrogen gas is detectable in dark galaxy candidates, even with the absence of optical counterparts \citep[e.g.][]{2021MNRAS.507.2905W,2022ApJ...926..167J,2024MNRAS.528.4010O}. In this regard, there have been several dark galaxy candidates suggested, which include VIRGOHI \citep{2005ApJ...622L..21M}, Dragonfly 44 \citep{2016ApJ...828L...6V}, AGESVC1 282 \citep{2020A&A...642L..10B}, AGC 229101 \citep{2021AJ....162..274L}, and FAST J0139+4328 \citep{2023ApJ...944L..40X}. It is still debated whether or not these {\HI} gases have cosmological origin. There are other possible explanations for the detection of {\HI} gas without optical counterparts; e.g. the intergalactic gas clouds that have experienced either ram pressure or tidal interactions with other galaxies \citep[e.g.][]{2015A&A...584A.113L,2017IAUS..321..238L}.
Therefore, it is necessary to examine the physical properties of dark galaxies with a larger sample to better understand the formation mechanism. For this purpose, we systematically identify dark galaxy candidates using the ALFALFA untargeted {\HI} survey in this study \citep{2005AJ....130.2598G,2011AJ....142..170H,2018ApJ...861...49H}. We carefully examine their properties using both internal and external physical parameters to understand what makes them unique.

This paper is organized as follows. In Section ~\ref{sec:data}, we begin by describing our selection criteria for dark galaxy candidates in the ALFALFA sample. We then explain the details regarding the computation of galaxy parameters.
In Section~\ref{sec:results}, we present the catalog of dark galaxy candidates.
In Section~\ref{sec:discussion}, we examine the physical properties of the candidates and compare them to those of luminous galaxies in the ALFALFA and SDSS DR7 samples.
Finally, in Section~\ref{sec:conclusion} we summarize our results and suggest possible future work.

Throughout, we adopt the flat $\rm \Lambda CDM$ model with $\Omega_{m}=0.3$, $\Omega_{\Lambda}=0.7$, and $ H_0 = 70\,\mathrm{km}\,\mathrm{s}^{-1}\,\mathrm{Mpc}^{-1}$.

\section{data}\label{sec:data}
\subsection{The ALFALFA Survey}
The Arecibo Legacy Fast ALFA (ALFALFA) survey \citep{2005AJ....130.2598G, 2011AJ....142..170H} mapped over $\sim 7000 \, \rm deg^2$ of the high galactic latitude sky to detect extragalactic {\HI} sources, exploiting the large collecting area of the Arecibo telescope. The survey covered the Declination range of 0 to 36 degrees. The two disconnected survey regions, designated as the ALFALFA ``spring" and ``fall" regions, encompass $\rm 07^h30^m \, < \, R.A. \, < \, 16^h 30^m$ and $\rm 22^h\, < \, R.A. \, < \, 3^h$, respectively.
The survey was conducted using the seven beam L-band $\rm (1.4\,GHz)$ receiver, which covers $1335-1445 \, \rm MHz$ with 4096 equally spaced spectral channels; this corresponds to a redshift range of $-2000 \, < \, cz_{\odot}\, (\rm km \, s^{-1}) \, < 17912 $ with a spectral resolution of $5.1 \rm \, km \, s^{-1}$ at $1420.4 \, \rm MHz$. The standard ALFALFA grid covers $2.4^\circ \times 2.4^\circ$ with the beam size of $3.8' \times 3.3'$ FWHM. They perform signal detection using a matched filter algorithm. To estimate the reliability of signal detection, they use both simulated galaxies and false signals from the ALFALFA data cubes within the velocity range of $-2000 \rm \, km\, s^{-1}$ to $-500 \rm \, km\, s^{-1}$, which is uncontaminated by Galactic and extragalactic {\HI} signals. The confidence level of detections with signal-to-noise ratio (SNR) of $> 6.5$ is $\sim 95\%$ \citep{2007AJ....133.2087S}.

\cite{2018ApJ...861...49H} presented 31502 extragalactic {\HI} sources from the ALFALFA survey. As this is a ``untargeted" survey, they were able to identify 344 intriguing sources that have no optical counterparts among $\sim 25000$ high-quality detections (i.e. $\rm SNR \gtrsim 6.5$, {\HI} code = 1 in their Table 2). 

\subsection{Selection criteria for dark galaxy candidates}\label{sec:2.2}

\subsubsection{Removal of HI clouds: clustering properties}

\begin{figure}[t]
    \centering
    \includegraphics[width=0.45\textwidth]{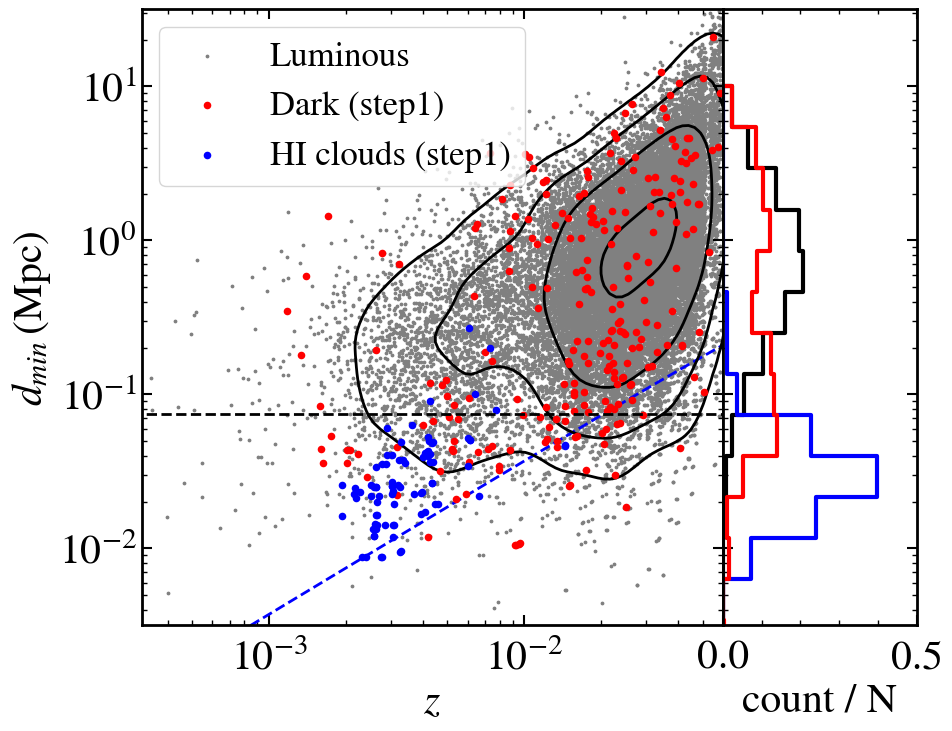}
    \caption{Selection criterion for the {\HI} clouds based on minimum projected distance. The colored dots represent ALFALFA sources without optical counterparts, whereas black dots (i.e. Luminous) represent ALFALFA high-quality detections with optical counterparts. Blue dots represent {\HI} clouds from the literature (step1) and red dots represent the remaining {\HI} sources without optical counterparts. The black dashed line denotes $d_{min}=0.075 \ \rm Mpc$, and the blue dashed line denotes Arecibo $3'$ beam size as a function of redshift}.
    \label{fig1}
\end{figure}

Gravitationally unstable {\HI} clumps, potentially formed through galaxy interactions, can also be detected as {\HI} sources without optical counterparts. To exclude these sources, we begin by reviewing the literature of extended source catalogs that include interacting pairs and small groups \citep{2019AJ....157..194H}. Among the 344 sources without optical counterparts, we find that 83 of them are likely tidally disrupted {\HI} clouds: e.g. Leo region \citep{1989ApJ...343...94S,2009AJ....138..338S,2016MNRAS.463.1692L}, NGC 4532/DDO 137 pair \citep{1999AJ....117..811H, 2008IAUS..244..362K}, NGC 4254 \citep{2007ApJ...665L..19H}, NGC 7448 \citep{2011MNRAS.415.1883D}, NGC 7769 \citep{1997AJ....114...77N}, and HI1225+01 \citep{1995AJ....109.2415C}. Here, {\HI} clouds refer to {\HI} sources with distinctive clumpy features that likely originate from tidal interactions, distinguishing them from primary {\HI} sources associated with host galaxies. Moreover, follow-up studies found that 13 of these {\HI} sources are likely to have optical counterparts \citep{2014MNRAS.443.3601L, 2014ApJ...787L...1C, 2015AJ....149...72C, 2015ApJ...801...96J, 2024ApJ...964...85D}. We therefore exclude these {\HI} sources and are left with 248 {\HI} sources for further analysis.

Although the known {\HI} clouds identified in the literature have been excluded, some may still remain in the sample. To further remove these clouds from our sample, we apply the projected distance to the nearest {\HI} source ($d_{min}$) criterion. We compute the projected distance to the nearest {\HI} source for a target {\HI} source as follows: $d_{min}=\chi\theta_{min} / (1+z)$, where $\chi$ is the comoving distance to the sample, and $\theta_{min}$ is the angular separation between the nearest ALFALFA source in radial velocity range of $400 \, \rm km \, s^{-1}$ from the target source. This selection criteria arises from the fact that most {\HI} clouds in the literature (i.e. $\gtrsim95\%$) have interacting pairs with a radial velocity difference smaller than $400\, \rm km \, s^{-1}$. Figure~\ref{fig1} shows the distributions of the projected distance to the nearest {\HI} source as a function of redshift. This figure shows that the known {\HI} clouds from the literature tend to be close to other {\HI} sources (i.e. small $d_{min}$). We therefore exclude 56 sources that have $d_{min}< 0.075 \, \rm Mpc$. It should be noted that most {\HI} sources in Figure~\ref{fig1} do not have neighboring sources with the projected distance smaller than the Arecibo $3'$ beam size (i.e. blue dashed line) and our selection criterion of $d_{min}=0.075\, \rm Mpc$ is conservative enough to remove the {\HI} sources with pair separation smaller than the Arecibo $3'$ beam size. However, to be more strict, we remove additional 10 sources that have another ALFALFA source within the $3'$ ALFALFA beam regardless of optical counterparts.

\subsubsection{Removal of HI sources using NASA/IPAC Extragalactic Database and Dark Energy Survey Instrument Survey Data}

To further refine our dark galaxy candidates, we search the NASA/IPAC Extragalactic Database (NED) and Data Release 1 of the Dark Energy Survey Instrument (DESI) survey catalog \citep{2025arXiv250314745D} for potential optical counterparts or associated galaxies corresponding to the remaining {\HI} sources. We first compute the distribution of angular separations between {\HI} source and optical counterparts as a function of SNR in the ALFALFA sample. We then exclude {\HI} sources that satisfy the following; a galaxy is considered a potential counterpart if the velocity difference between NED or DESI and ALFALFA is smaller than $2\times W_{50}$ of the {\HI} source AND the angular separation between the two is smaller than $5\sigma$ of the ALFALFA positional uncertainty. Furthermore, we find that most {\HI} clouds from the literature have interacting pairs with a radial velocity difference smaller than $400 \, \rm km \, s^{-1}$ and a projected distance smaller than $0.075 \,\rm Mpc$.
We apply these conditions to exclude {\HI} sources that may have originated from tidal interactions with associated galaxies; $c\Delta z < 400 \, \rm km \, s^{-1}$ and $\Delta \theta < \text{max}\{ 5\sigma,\, 0.075 \, \rm Mpc / Distance \}$. In this way, we remove 24 sources with potential optical counterparts, and 16 sources possibly originated from tidal interactions. The remaining 142 {\HI} sources comprise our final sample of dark galaxy candidates.

\subsubsection{Examination of DESI Legacy Survey and Multi-wavelength Imaging Data}
\begin{figure*}[t]
    \centering    
    \includegraphics[width=0.75\textwidth]{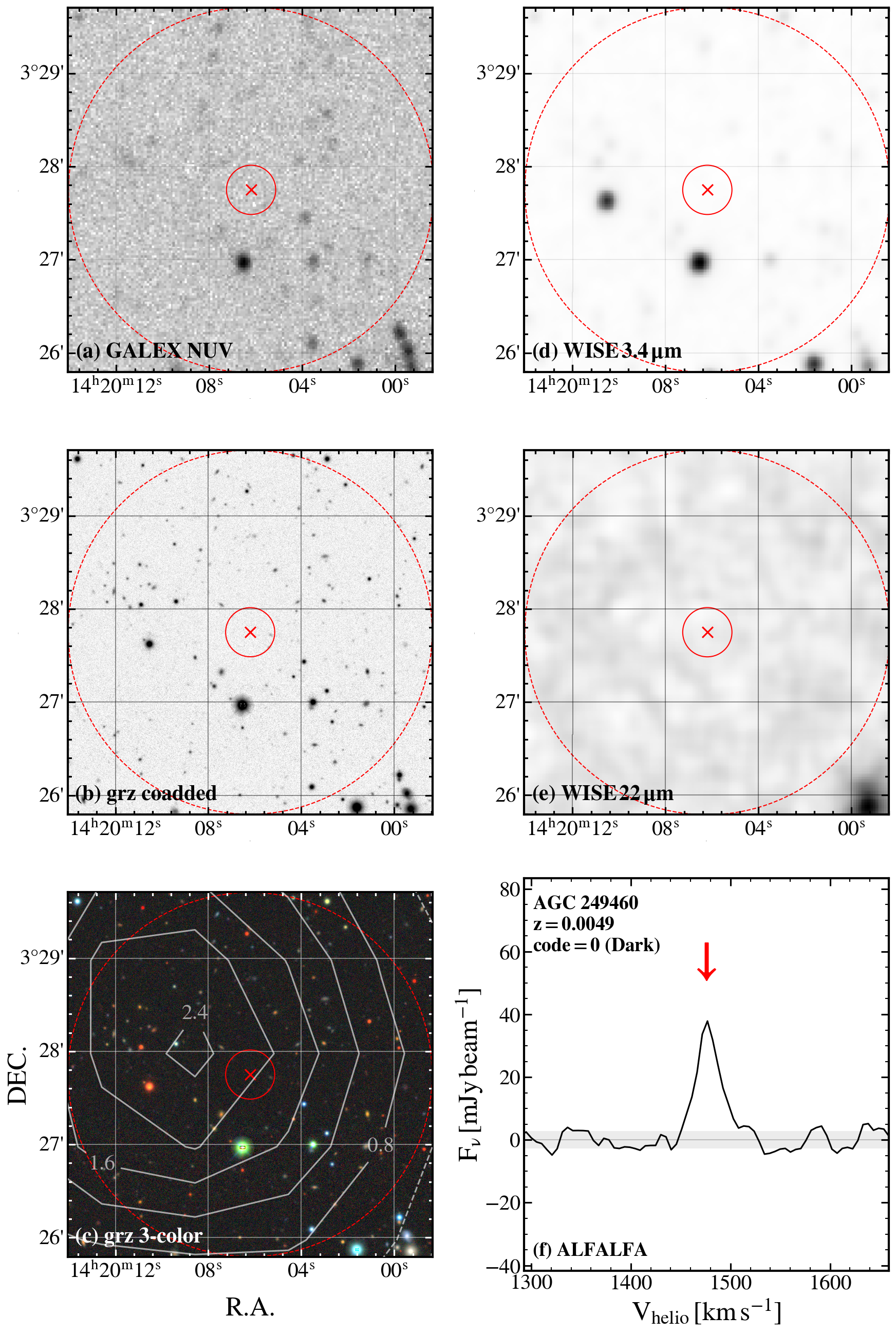}
    \caption{Example cutout images for a dark galaxy candidate AGC 249460. (a) The GALEX NUV (2315$\, \rm \AA$). (b) The DESI Legacy Imaging Survey $g$, $r$, $z$ coadded image. (c) The DESI Legacy Imaging Survey $g$, $r$, $z$ 3-color composite image. White contours are the {\HI} intensity within $\rm 2\times W_{50}$ velocity width from the ALFALFA data cubes, with the lowest contour (dashed line) corresponding to the $2\sigma$ level. The labels on the contours indicate the {\HI} column density $N_{\rm  HI} \rm \,[10^{19} cm^{-2}]$. (d) The WISE band 1 (3.4$\mu m$). (e) The WISE band 4 (22$\mu m$). (f) The ALFALFA {\HI} spectrum from \citet{2018ApJ...861...49H}. The red arrow represents the central heliocentric velocity of the {\HI} source and the shaded region represents the $1\sigma$ noise level. In all photometric images, the outer red dotted lines and the inner red solid lines denote 5$\sigma$ and median angular separation between the optical counterpart and the {\HI} source in the ALFALFA survey. The complete figure set (344 images) is available in the online journal.}
    \label{fig:2}
\end{figure*}

We utilized deep optical imaging to quantitatively determine the probability that a {\HI} source has an optical counterpart. To do that, we use the optical images from Data Release 10 of the DESI Legacy Imaging Survey \citep{2019AJ....157..168D}. The survey consists of three projects including DECaLs, BASS and MzLS, and covers $\sim 14000 \ \rm deg^2$ of the northern sky in three optical bands ($g\, , r\, ,z$). The ALFALFA footprint generally overlaps with the DECaLs survey, resulting in an $r$-band surface brightness limit of $28.5 \, \rm mag \, arcsec^{-2}$ ($3\sigma$ with $10''\times10''$ boxes, see \citealt{2023A&A...671A.141M} for details). The $5\sigma$ depths for the fiducial DESI target (i.e. galaxy with an exponential disk profile with a half-light radius of $0.45 \, \rm arcsec$) are $g=24.0$, $r=23.4$ and $z=22.5$ in AB magnitudes. 
We use coadded $g$, $r$, $z$ images and 3‑color composite images to identify potential optical counterparts.
We also use data from the Galaxy Evolution Explorer (GALEX) in NUV channel (2315$\rm \AA$) \citep{2007ApJS..173..682M}, the Wide-field Infrared Survey Explorer (WISE) in band 1 $(3.4\mu m)$ and 4 $(22 \mu m )$ \citep{2010AJ....140.1868W}, and from AKARI in band S $(90 \mu m)$ \citep{2015PASJ...67...50D} to identify potential ultraviolet/infrared emission from stellar or dust components. We construct cutout images for each galaxy to determine whether there is an optical counterpart, as shown in Figure~\ref{fig:2}; we provide comments regarding possible optical counterparts found from the cutout images in Table~\ref{tab:1} (and we further discuss this in Section~\ref{sec:results}). Figure~\ref{fig:2} for all the candidates are available as a figure set. It should be noted that the AKARI $90\mu m$ cutouts are not included in Figure~\ref{fig:2} because they do not show any clear emission around the center even though we have performed the visual inspection.

\subsection{Physical properties of dark galaxy candidates}
\subsubsection{Likelihood ratio of the candidates using optical imaging data}
\begin{figure}[t]
    \centering
    \includegraphics[width=0.45\textwidth]{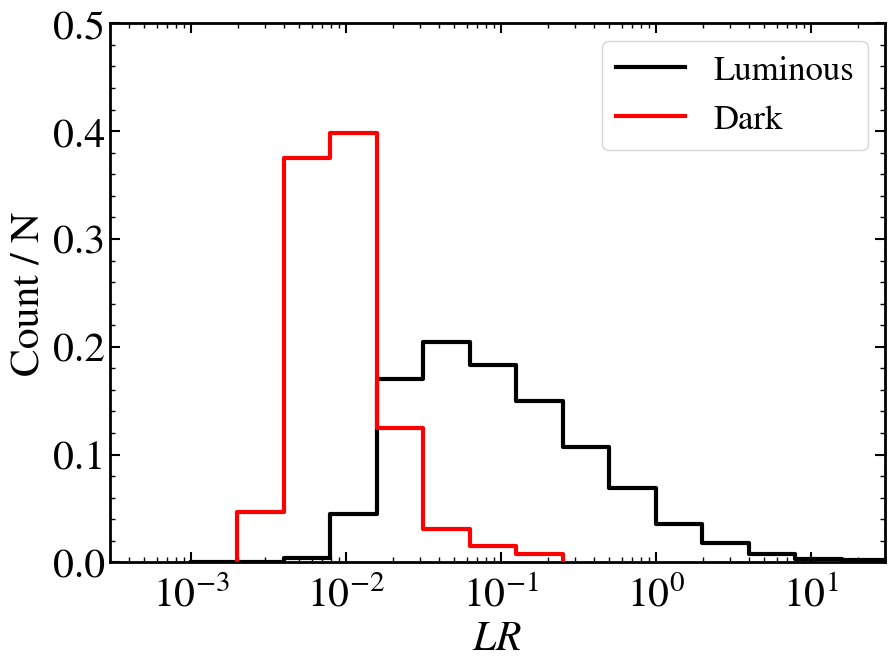}
    \caption{The distributions of likelihood ratio for dark galaxy candidates (red) and luminous galaxies (black).}
    \label{fig:add1}
\end{figure}

Although we exclude galaxies with optical counterparts using the selection criteria described in previous sections and examine the imaging data, there could be still luminous galaxies in the sample that are not detected due to their low surface brightness. To quantify the probability that an {\HI} source has an optical counterpart, we compute the likelihood ratio \citep{1992MNRAS.259..413S} for each {\HI} source as follows. The likelihood ratio of an {\HI} source is defined as

\begin{equation}
    LR=\sum_{\tilde d_i<5}\frac{Q_i\exp \left(-\tilde d_i^2/2\right)}{2\pi \sigma^2 n} , 
    \label{eq:1}
\end{equation}
where $n$ is the surface number density of sources with an $r$-band magnitude $m_r < 23.5$. The normalized projected distance $\tilde d$ is given by $\tilde d = d / \sigma$, with $d$ being the projected separation to a nearby source and $\sigma$ being the uncertainty of {\HI} source position. The factor $Q$ serves as a prior probability based on source properties. For simplicity, we set $Q$ equal to the $r$-band flux (in the unit of nanomaggies) for extended sources and equal to zero for point sources. Assuming that any nearby sources could possibly be associated with the {\HI} source, we sum the likelihood ratio over all sources with $\tilde d < 5$. Here, a higher likelihood ratio implies a higher probability that the {\HI} source has an optical counterpart. We compute the likelihood for each {\HI} source using the photometric catalog of extended sources from the $r$-band images of the Legacy Survey DR10. Most luminous galaxies (i.e. $\gtrsim 95\%$) have likelihood ratios larger than 0.016, whereas about $80\%$ of the dark galaxy candidates have likelihood ratios smaller than 0.016 (see Figure~\ref{fig:add1}). It should be noted that we do not exclude the dark galaxy candidates with likelihood ratios larger than 0.016 because those are the cases where nearby optical sources have no redshift information or the redshifts of optical sources differ from those of {\HI} sources.

\subsubsection{Local environments of dark galaxy candidates}\label{sec:2.3.2}

To study the environments of our dark galaxy candidates, we compute large-scale environmental parameters using the spectroscopic sample of galaxies from the SDSS main survey \citep{2009ApJS..182..543A}. The spectroscopic completeness of this sample is poor for bright galaxies because of saturation and cross-talk in the spectrograph, and for galaxies in high-density regions because of fiber collisions. We therefore use the SDSS data supplemented with redshifts from the literature for the galaxies with $m_r  <  17.77 \rm \, mag$ (see \citealt{2010A&A...522A..33H} for details).

To mitigate the effect of the redshift distortion caused by peculiar motion along the line of sight on the computation of three-dimensional environmental parameters, we use a method similar to \cite{2004ApJ...606..702T} and \cite{2016ApJ...818..173H} \citep[see also][]{2020MNRAS.491.4294K}. To do that, we first compute the mean separation $(d_{mean})$ to the nearest galaxy; the nearest galaxy is found in 3-D redshift space contracted by a factor of five along the line of sight direction compared to the perpendicular direction. This results in $d_{mean} = 1$ and 2.5 $\rm cMpc$ at redshifts $z=0.01$ and $0.06$, respectively. We then run the friends-of-friends algorithm \citep{1982ApJ...257..423H} using the linking lengths of $0.2\times d_{mean}$ perpendicular to the line of sight and $d_{mean}$ along the line of sight. For the structures identified in this way, we contract the line-of-sight velocities to have the velocity dispersion the same as the one perpendicular to the line-of-sight.

\begin{figure}[t]
    \centering
    \includegraphics[width=0.45\textwidth]{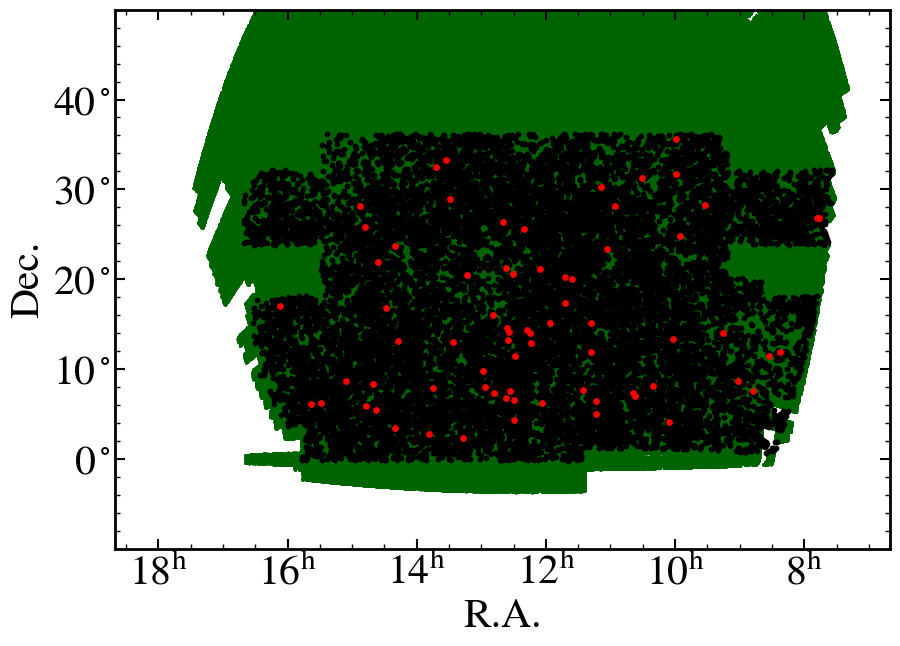}
    \caption{The sky coverage for the computation of local density. The green region represents the SDSS DR7 main survey area, while the black region represents the area covered by the ALFALFA sources. The red dots are the dark galaxy candidates within the SDSS DR7 main survey area.}
    \label{fig:4}
\end{figure}

We use this update sample of SDSS galaxies to compute the local density of dark galaxy candidates and luminous galaxies with {\HI} detection. To do that, we first make a volume-limited sample of SDSS galaxies using the criterion of evolution and K-corrected $r$-band magnitude within $M_r < -19.2$. Then, the local density at a given position $\mathbf{x}$ is defined as
\begin{equation}
\rho_{n}(\mathbf{x})=\sum_{i=1}^{n} \gamma_i L_i W(|\mathbf{x}_i-\mathbf{x}|,h),
\label{eq:2}
\end{equation}
where $n$ is the number of neighboring galaxies used for the local density estimation, $\gamma$ is the mass-to-light ratio, $L$ is the evolution and K-corrected $r$-band luminosity. We also use the standard Smoothed-Particle Hydrodynamics (SPH) kernel $W$ in the computation as in \cite{2007ApJ...658..898P} and \cite{2012MNRAS.419.2670M}. The smoothing length $h$ is set to the distance to the $n$-th nearest galaxy. We compute the local density using 20 nearby galaxies, $n=20$, because the distance to the 20th nearest galaxy corresponds to a scale of a few Mpc, which is suitable for studying the impact of the local environment on galaxy properties \citep{2007ApJ...658..898P}. Furthermore, we choose a mass-to-light ratio of $\gamma(\rm early)=2\gamma(\rm late)$ as in \cite{2008ApJ...674..784P} because the central velocity dispersion of early-types is approximately $\sqrt2$ times that of late-types in the range of $M_r<-19.5$.
The mean mass density within the survey volume is then defined as
\begin{equation}
\bar{\rho} = \sum_{\text{all}} \frac{\gamma_i L_i}{V}.
\end{equation}
To avoid the boundary effect of the SDSS survey, we compute the local densities only for galaxies far from the survey boundary (i.e. $1 \, \rm deg$) as shown in Figure~\ref{fig:4}.

\subsection{The IllustrisTNG Cosmological Simulations}
To compare the physical properties of dark galaxy candidates in this study with those in simulations, we use the data from the IllustrisTNG simulation \citep{2018MNRAS.475..624N,2018MNRAS.477.1206N,2018MNRAS.480.5113M,2018MNRAS.475..676S,2018MNRAS.475..648P}. The IllustrisTNG (hereafter TNG) is a set of large-volume magneto-hydrodynamical cosmological simulations based on the flat $\rm \rm \Lambda CDM$ cosmological parameters from \cite{2016A&A...594A..13P}: $\Omega_\Lambda = 0.6911$, $\Omega_m = 0.3089$, $\Omega_b = 0.0486$, $\sigma_8 = 0.8159$, $n_s = 0.9667$, and $H_0 = 67.74\, \rm km \, s^{-1} \, Mpc $. Utilizing the moving-mesh code AREPO \citep{2010MNRAS.401..791S,2020ApJS..248...32W}, the TNG simulations incorporate self-gravity, hydrodynamics, and cosmic magnetic fields. They also employ advanced sub-resolution models accounting for processes such as gas cooling, star formation, metal enrichment, and feedback from supernova (SN) and active galactic nuclei (AGN). The TNG simulations consist of three realizations: TNG50, TNG100, and TNG300, with each number representing the side length of the simulation box in $\rm cMpc$.

In this work, we use TNG50 as done for \cite{2024ApJ...962..129L} because it provides the highest mass resolution down to $8.5\times 10^4 \, \rm M_\odot$ for baryons \citep{2019MNRAS.490.3196P,2019MNRAS.490.3234N}, allowing us to identify low-mass {\HI}-detected galaxies in the simulation. We consider the SUBFIND groups as galaxies to construct a mock galaxy catalog similar to the ALFALFA sample. Here, the SUBFIND algorithm \citep{2001MNRAS.328..726S,2009MNRAS.399..497D} identifies gravitationally self-bound substructures within friends-of-friends groups using merger trees. Because the TNG simulations do not model gas temperature below $10^4 \, \rm K$, the amount of atomic hydrogen gas ({\HI}) in each galaxy is modeled in post-processing \citep{2018ApJS..238...33D,2019MNRAS.487.1529D}. We adopt mass of the atomic hydrogen for each SUBFIND group from the model of \cite{2014ApJ...790...10S} among various models for the atomic and molecular hydrogen transition. It should be noted that there are subtle differences in the gas properties of simulated galaxies depending on the resolution of the simulation \citep[e.g. TNG100 vs. TNG300 as in][]{2018ApJS..238...33D,2019MNRAS.487.1529D}; this is also true for stellar properties as in \citet{2018MNRAS.475..648P}. Therefore, the quantitative measures of physical properties of simulated galaxies to be compared with those of observed galaxies may not be perfectly matched.

\begin{figure}[t]
    \centering
    \includegraphics[width=0.45\textwidth]{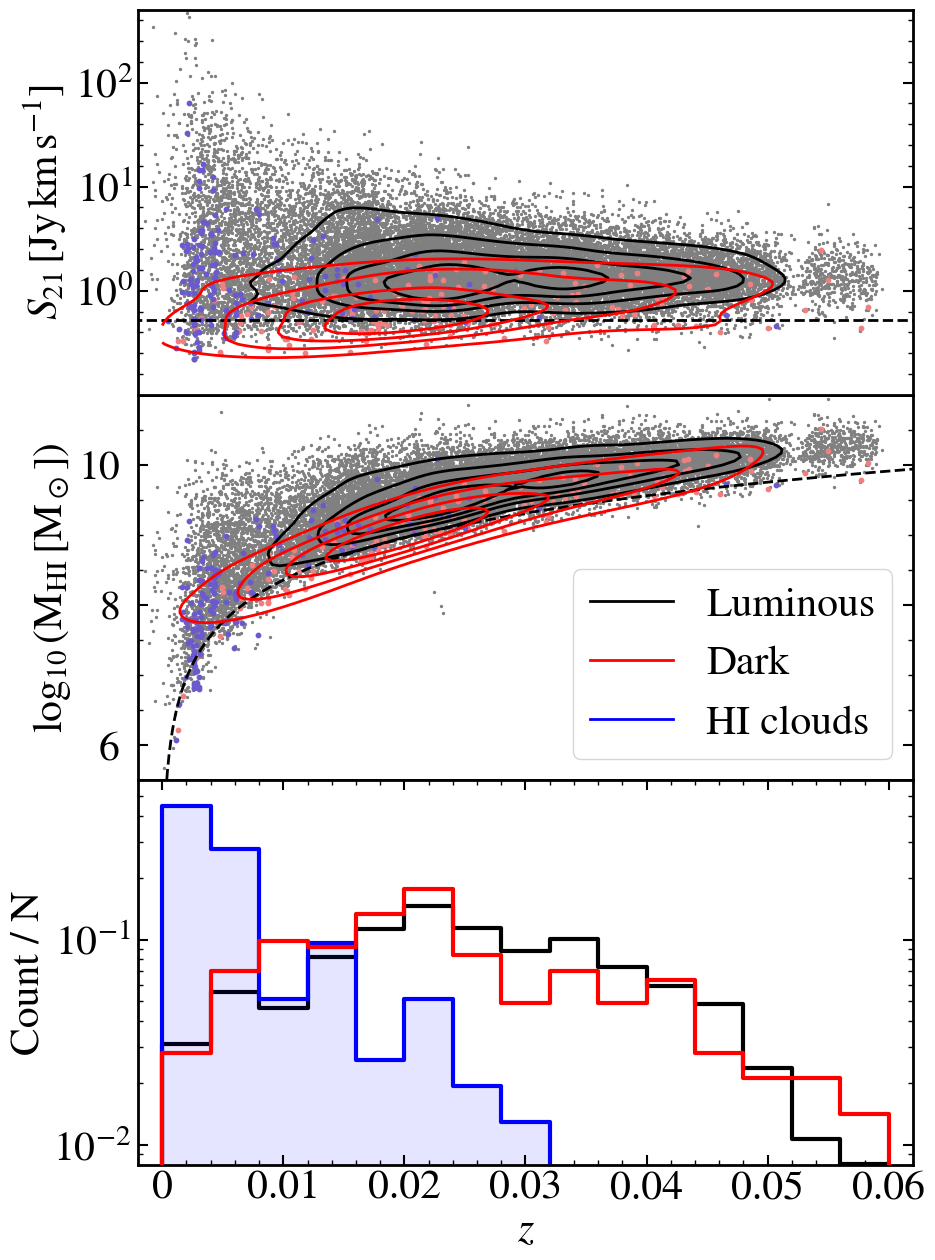}
    \caption{The distributions of integrated {\HI} flux and {\HI} mass for the dark galaxy candidates, luminous galaxies, and {\HI} clouds, as a function of redshift. The dashed lines in the first and second panels represent $\rm SNR=5$, assuming $W_{50} = 100\ \rm{km\,s^{-1}}$ and median RMS (i.e. $\sigma_{\rm rms}=2.34\, \rm mJy$) of the ALFALFA spectra.}
    \label{fig:5}
\end{figure}
We construct an ALFALFA-like galaxy catalog from the TNG50. We first assign a random redshift ($z<0.06$) to each galaxy, where the probability of a galaxy having a specific redshift is proportional to the volume at that redshift; this is because the volume of the TNG50 is much smaller than that of the ALFALFA survey. Next, we identify galaxies with {\HI} masses larger than the $5\sigma$ detection limit
(i.e. $\rm \log_{10}\left( M_{HI} \, [M_{\odot}]\right) > 8.3$ at $z=0.01$; see the dashed line in the middle panel of Figure~\ref{fig:5}) as TNG ALFALFA-like galaxies. If the SDSS $r$-band magnitude of a galaxy computed from stellar particles is fainter than $m_r=17.77$, we further classify it as a TNG dark galaxy candidate. As a result, we are left with $\sim 880$ TNG ALFALFA-like galaxies, of which $\sim 80$ are dark galaxy candidates and the remaining $\sim 800$ are luminous galaxies. As the selection criteria for dark galaxy candidates in simulations are not exactly the same as those in observations and {\HI} modeling in simulations may not be perfect, the comparison between simulations and observations in Section~\ref{sec:discussion} should be considered in a qualitative sense.

\section{Results}\label{sec:results}
From the list of 344 extragalactic {\HI} sources without optical counterparts \citep{2018ApJ...861...49H}, we identified 142 dark galaxy candidates following the selection criteria described in Section~\ref{sec:2.2}. Table~\ref{tab:1} lists all 344 initial candidates with the original 10 columns \citep[see Table 2 in][]{2018ApJ...861...49H} and 7 additional columns from this study. The description of the additional columns are as follows.

\begin{table*}
    \renewcommand\thetable{2}
    \centering
    \caption{The quality code for dark galaxy candidates}
    \begin{tabular*}{0.9\textwidth}{cc@{\extracolsep{0.05\textwidth}}c}
    \hline
    \hline
    Code& N & Description\\
    \hline
    0 & 142 & Final dark galaxy candidates\\
    1 & 83  & {\HI} clouds identified in the literature\\
    2 & 13  & {\HI} sources with optical counterparts identified in the literature\\
    3 & 56  & {\HI} clouds excluded by the minimum projected distance criterion\\
    4 & 10  & {\HI} sources with another {\HI} source within the ALFALFA beam\\    
    5 & 16  & {\HI} sources with associated galaxies identified using the NED and DESI catalog\\
    6 & 24  & {\HI} sources with potential optical counterparts identified using the NED and DESI catalog\\

    \hline
    \label{tab:2}
    \end{tabular*}
\end{table*}
\begin{itemize}
    \item column (11) - The logarithmic dynamical mass and its uncertainty in solar mass units. We estimate the dynamical mass using the circular velocity from $W_{20}$ and the {\HI} radius. The {\HI} radius is calculated by converting the {\HI} mass, as the {\HI} size is tightly correlated with {\HI} mass within $\sigma \sim 0.06 \rm \ dex$ \citep[{\HI} size-mass relation]{2016MNRAS.460.2143W}. We estimate the uncertainty by propagating the uncertainties of the {\HI} mass, the {\HI} size-mass relation, and its scatter.
    \item column (12) - {\HI} line asymmetry. We compute the asymmetry using the {\HI} line profile from \citet{2018ApJ...861...49H}. The {\HI} line asymmetry is defined by
    \begin{equation}
        A= \frac{\left| F_{\rm left} -F_{\rm right}\right|}{ F_{\rm left} +F_{\rm right}},
    \end{equation}
    where $F_{\rm left}$ and $F_{\rm right}$ are the fluxes integrated over the radial velocity ranges $\rm (V_{helio} - W_{50},V_{helio})$ and $\rm (V_{helio},V_{helio}+W_{50})$, respectively.
    \item column (13) - The local density derived from eq.~\ref{eq:2}. We normalize the local density to the mean density within the ALFALFA survey volume. 
    \item column (14) - The likelihood ratio of the sample derived from eq.~\ref{eq:1}.
    \item column (15) - The minimum projected distance to other ALFALFA sources in units of Mpc.
    \item column (16) - A quality code for the reliability of dark galaxy candidates. The catalog includes all extragalactic {\HI} sources without optical counterparts in the ALFALFA survey. Table~\ref{tab:2} lists the quality code along with the corresponding description for the dark galaxy candidates.
    \item column (17) - Note for {\HI} sources, such as potential optical counterparts, related galaxies, or existence of foreground stars.
\end{itemize}

\section{Discussion}\label{sec:discussion}
\begin{figure*}[t!]
    \centering
    \includegraphics[width=0.95\textwidth]{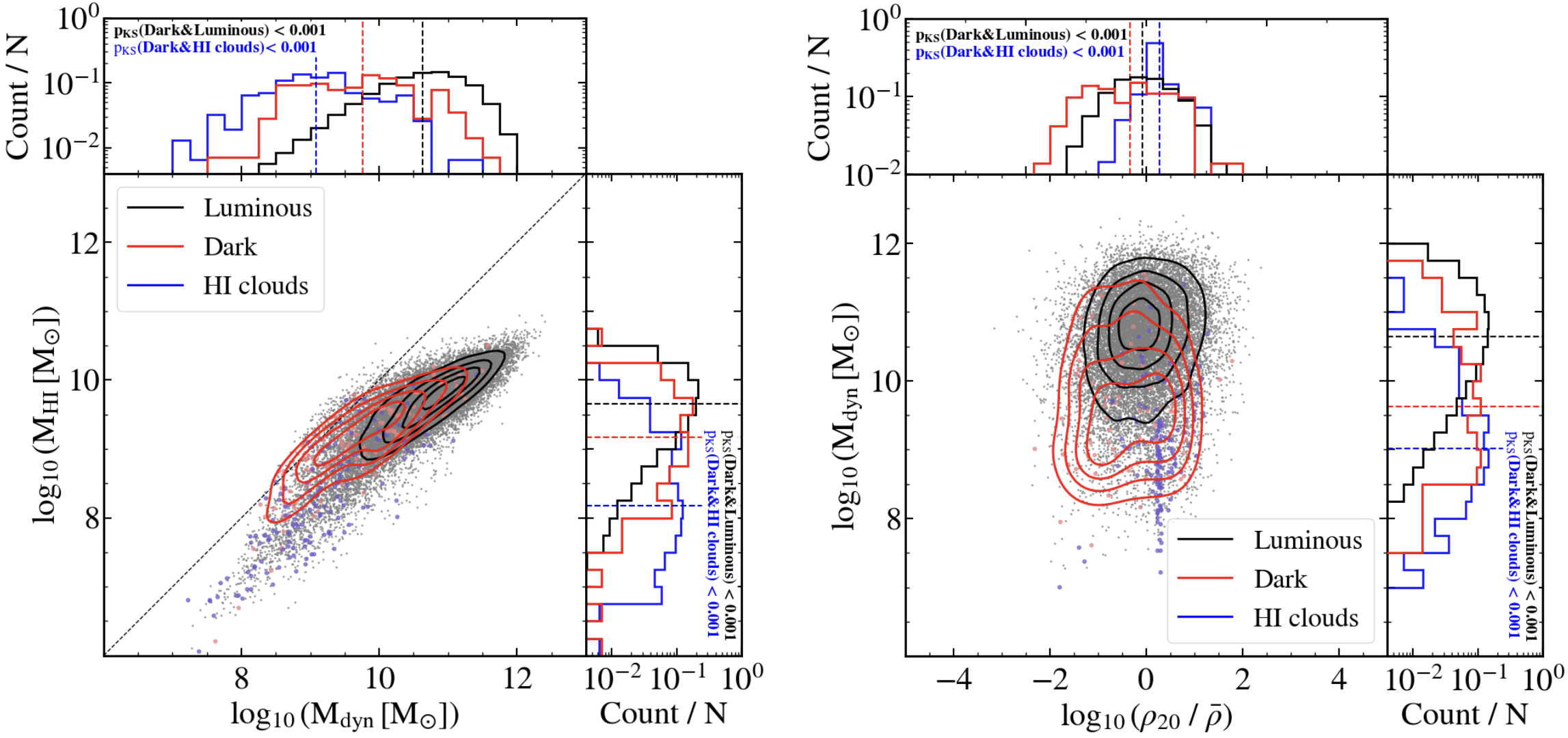} \\
    \vspace{5mm}
    \includegraphics[width=0.95\textwidth]{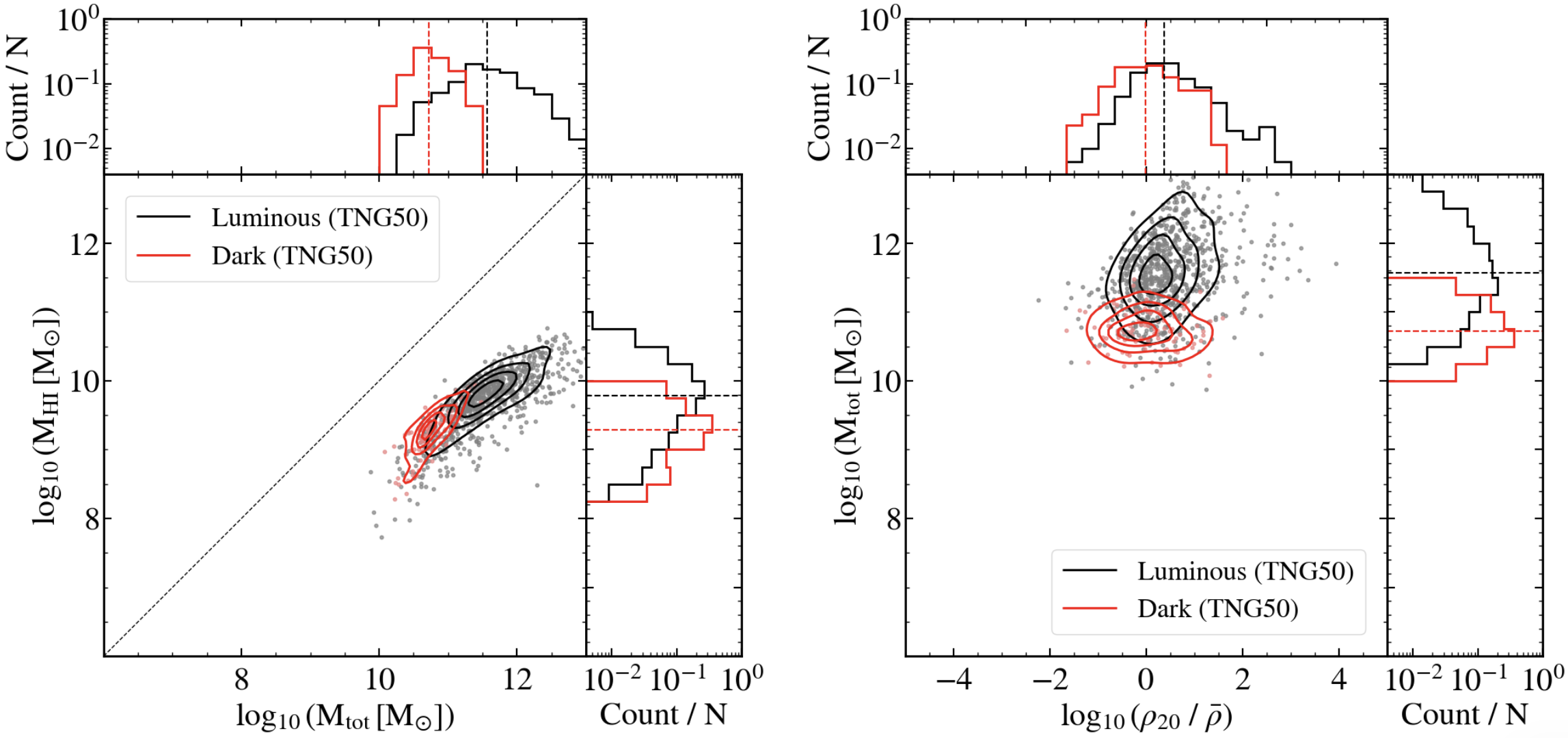}
    \caption{The internal properties and local environments of the ALFALFA galaxies (\textit{top}) and the TNG50 simulation (\textit{bottom}). The black, red and blue symbols represent the luminous, dark galaxy candidates and {\HI} clouds, respectively. \textit{left}: {\HI} and dynamical(total) mass distribution of the ALFALFA(TNG50) galaxies. \textit{right}: The local density and dynamical(total) mass distribution of the ALFALFA(TNG50) galaxies. The p-values from the Kolmogorov-Smirnov test are shown in each histogram panel, indicating the statistical significance of the differences between the distributions of dark galaxy candidates and other samples.}
    \label{fig:6}
\end{figure*}

\begin{figure*}[t]
    \centering
    \includegraphics[width=0.8\textwidth]{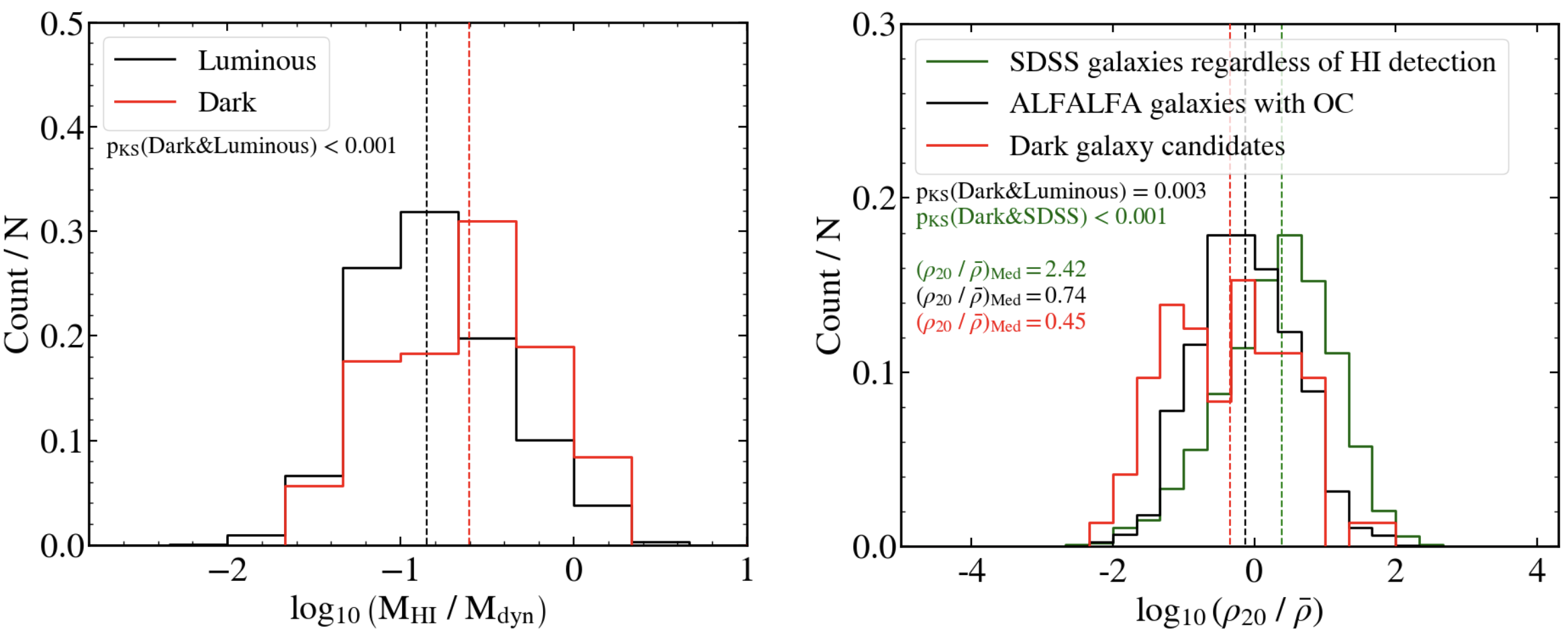 }
    \caption{The distributions of {\HI}-to-dynamical mass ratio and local density for the dark galaxy candidates and the comparison samples. In both panels, the red lines represent the dark galaxy candidates. In the left panel, the black histogram represents the mass controlled luminous galaxies in the ALFALFA sample. In the right panel, the black and green histograms represent luminous galaxies from the ALFALFA sample and SDSS DR7 galaxies, respectively; their mass and redshift distributions are matched to those of the dark galaxy candidates.}
    \label{fig:7}
\end{figure*}

In this section, we compare the physical properties of dark galaxy candidates with those of luminous galaxies to find any distinctive features between the two.

\subsection{Dynamical and {\HI} gas masses of dark galaxy candidates}
To study the internal properties of our candidates, we first examine their dynamical and {\HI} gas masses. We compute the {\HI} masses using their distances and integrated {\HI} line fluxes from the ALFALFA survey. We calculate the dynamical mass using the rotational velocity derived from $W_{20}$, the line width measured at the $20\%$ level of the peak (column 8 in Table~\ref{tab:1}), and the {\HI} radius obtained from the {\HI} size-mass relation \citep{2016MNRAS.460.2143W}. The top left panel of Figure~\ref{fig:6} shows the mass distribution of dark and luminous galaxies in the ALFALFA sample. Here, the luminous galaxies refer to $\sim 31000$ extragalactic {\HI} sources with optical counterparts from \cite{2018ApJ...861...49H}. 

The overall distribution of the contours shows that the dark galaxy candidates are generally less massive than luminous galaxies. Although we do not apply any {\HI} flux selection criteria to either the dark or luminous galaxies (see Figure~\ref{fig:5} for the distributions of {\HI} flux and {\HI} gas mass as a function of redshift for both samples), the {\HI} gas mass distribution of the dark galaxy candidates (red contours) extends toward lower masses compared to luminous galaxies. The median {\HI} gas mass of the dark galaxy candidates is about one-seventh that of the luminous galaxies. The difference between dark galaxy candidates and luminous galaxies is even more prominent in the dynamical mass distributions, suggesting that dark galaxy candidates possess higher {\HI}-to-dynamical mass ratios than luminous galaxies. The Kolmogorov–Smirnov (K-S) test to examine whether the distributions are drawn from the same parent distribution suggests that the statistical significance of the difference in {\HI} and dynamical masses is larger than $3\sigma$ (see the p-values in the histogram panel).
The bottom left panel of Figure~\ref{fig:6} shows the {\HI} and total mass distributions of ALFALFA-like luminous galaxies and dark galaxy candidates in the TNG50 simulation. We find an overall tendency for dark galaxy candidates to be less massive and to have higher {\HI}-to-dynamical mass ratios compared to luminous galaxies in the simulation. This trend is consistent with our observational results. It should be noted that selection criteria for the dark galaxy candidates and the physical properties (e.g. total mass and {\HI} mass) in the simulation do not exactly match those in the observations; this is mainly due to limitations of current cosmological hydrodynamical simulations in terms of mass/spatial resolution, volume, and subgrid physics \citep[e.g.][]{2023ARA&A..61..473C}. Therefore, comparisons between simulations and observations in this work should be considered only in a qualitative sense.

When we compare {\HI}-to-dynamical mass ratios between luminous galaxies and dark galaxy candidates, we need to match the mass distribution for a fair comparison because {\HI} properties are closely correlated with its total/dynamical mass \citep{2015MNRAS.452.2479B}. In this regard, we construct a subsample of ALFALFA galaxies with optical counterparts, which has a dynamical mass distribution similar to that of the dark galaxy candidates to compare their {\HI}-to-dynamical mass ratios. The left panel of Figure~\ref{fig:7} shows the distributions of the {\HI}-to-dynamical mass ratios for the dark galaxy candidates and the luminous galaxies after the dynamical mass distribution is matched. After we remove the mass bias, dark galaxy candidates still generally show higher {\HI}-to-dynamical mass ratios than luminous galaxies.

The existence of dark galaxy candidates with low masses and high {\HI}-to-dynamical mass ratios appears consistent with the result from numerical simulations. The less massive halos could make them vulnerable to heating from cosmic reionization, which differentiates their gas properties from those of luminous galaxies. Their gas cannot efficiently form stars and remains mostly neutral, whereas luminous galaxies can form stars and ionize their gas \citep{2017MNRAS.465.3913B,2024ApJ...962..129L}.

\subsection{How special is the environment for dark galaxy candidates?}

\begin{figure*}[t]
    \centering
    \includegraphics[width=0.8\textwidth]{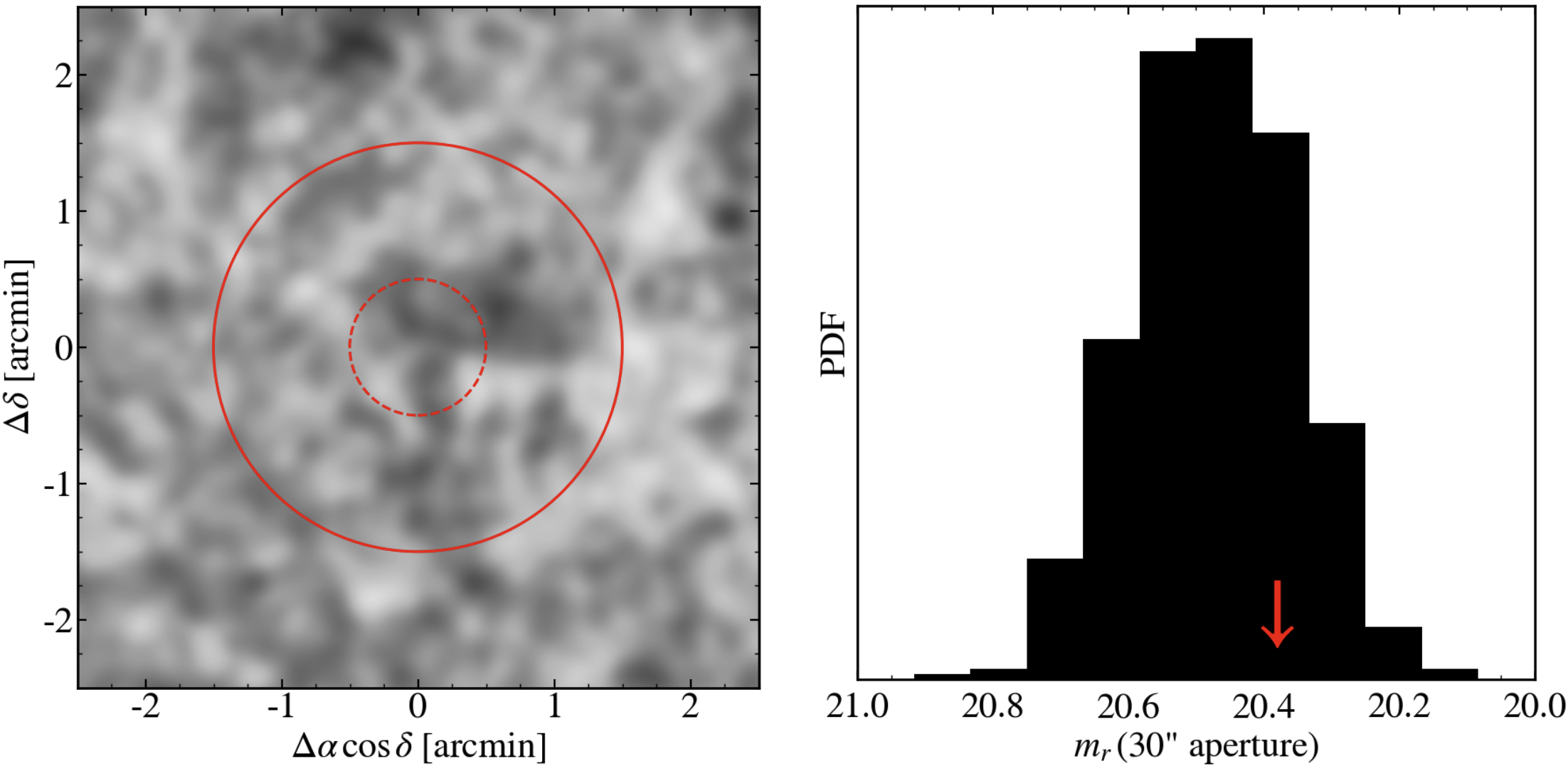} 
    \caption{($\it left$) Median-stacked $r$-band image of the dark galaxy candidates, smoothed with a Gaussian kernel of $\sigma=5''$. The red solid and dashed circle represent radii of $1.5'$ and $0.5'$, respectively. ($\it right$) Magnitude distribution of the median-stacked image, measured at random positions using a circular aperture with a radius of $30''$. The red arrow indicates the magnitude at the center.}
    \label{fig:8}
\end{figure*}

We show the distribution of local density as a function of dynamical mass for luminous galaxies and dark galaxy candidates along with {\HI} clouds in the top right panel of Figure~\ref{fig:6}. It is clear that their distributions are not similar, which is confirmed by the K-S test; the histogram of local density for the dark galaxy candidates is shifted toward less dense regions compared to luminous galaxies, which is consistent with the result from simulations (see Fig. 6 in \cite{2024ApJ...962..129L}).
We find the effect of the {\HI} selection criterion (i.e. the minimum projected distance to the nearest {\HI} source) on the local density distribution to be negligible.
It is also noted that the distribution of dark galaxy candidates differs from that of {\HI} clouds. The {\HI} clouds tend to avoid under-dense regions; the $\rho_{20}$ values of the {\HI} clouds are generally higher than those of dark galaxy candidates, even after we exclude their host galaxies from the calculation. This is consistent with the idea that the {\HI} clouds are mainly tidal debris, which generally form in over-dense regions \citep[e.g.][]{2007ApJ...665L..19H,2008IAUS..244..362K,2011MNRAS.415.1883D, 2016MNRAS.463.1692L}.
The bottom right panel of Figure~\ref{fig:6} shows the total mass and local density distributions of ALFALFA-like luminous galaxies and dark galaxy candidates in the TNG50 simulation. We find that dark galaxy candidates tend to reside in less dense regions compared to luminous galaxies in the simulation, which is again consistent with our observational result.

Although we construct the sample of dark galaxy candidates from an untargeted survey, there may still be a lingering mass selection bias because the intrinsic mass distribution of dark galaxy candidates differs from that of luminous galaxies. This bias should be mitigated when comparing the local density between dark and luminous galaxies because the environment of galaxies (e.g. local density) is strongly correlated with the internal properties of galaxies (e.g. mass) \citep{2007ApJ...658..898P}.
To mitigate this bias, we first construct a subsample of ALFALFA galaxies with optical counterparts to have a dynamical mass distribution and a redshift distribution similar to that of the dark galaxy candidates. Because ALFALFA galaxies (i.e. {\HI} detected galaxies) tend to be located in under-dense regions \citep[e.g.][]{2018ApJ...852..142C}, we construct another sample of galaxies that consists of SDSS galaxies regardless of {\HI} detection. Because most of these galaxies do not have {\HI} detections, we use stellar masses instead of dynamical masses to match the mass distributions between ALFALFA and SDSS galaxies. This procedure ensures that all three samples share similar mass and redshift distributions, which allows a fair comparison between local densities.

The right panel of Figure~\ref{fig:7} shows the local density histograms of the controlled samples. The local densities of the three samples are computed within the same sky coverage using the same tracers as described in Section~\ref{sec:2.3.2}. The SDSS galaxies regardless of {\HI} detection tend to be in denser regions compared to the other samples even though the extended feature toward very dense regions results partly due to the finger correction process. The dark galaxy candidates are located in less dense regions than the ALFALFA {\HI}-detected galaxies with optical counterparts. Moreover, there is a hint of bimodality in the local density histogram of the dark galaxy candidates, which could be confirmed by a larger sample in the future. The same panel also shows the median local density of each sample, indicating that the median local density of the dark galaxy candidates is about $40\%$ lower than that of the luminous galaxies.

These results agree with those from simulations, which show distinct properties for dark galaxies in the present epoch \citep[e.g.][]{2024ApJ...962..129L}. Compared to luminous galaxies, dark galaxies are more likely to be formed in less dense regions with little star-forming gas. Furthermore, they experience fewer mergers that could increase their mass and supply them with star-forming gas. \citep{1997MNRAS.292L...5J,2013MNRAS.432.1701K, 2020MNRAS.498L..93J,2024ApJ...962..129L}.

Dark galaxy candidates with higher masses or located in denser environments might have different origins. For example, as galaxies move across filamentary structures or fall into galaxy clusters, they experience ram pressure from the surrounding gas. This pressure could remove star-forming gas from the interior of the galaxies, leaving them dark \citep{1972ApJ...176....1G,2005A&A...437L..19O, 2018ApJ...856..160H}.

\subsection{Stacked image of dark galaxy candidates}

To examine the possible detection of our dark galaxy candidates in optical images, we perform a stacking analysis.
To do that, we extract $30' \times 30'$ cutouts in $r$-band from the Legacy Survey centered on each candidate. We then randomly flip the cutouts and construct a median-stacked image; we take the median value at each pixel by excluding values larger than 5 root mean square.

To estimate the statistical significance, we measure the flux at the center of the image within a circular aperture of radius $30''$ which corresponds to $\sim1\sigma$ positional uncertainty. We then repeat this measurement at random positions to estimate the background level and its fluctuations. As shown in the right panel of Figure~\ref{fig:8}, the central aperture yields a magnitude of 20.4, corresponding to $0.8 \sigma$ above the background. We also measure the flux with smaller apertures (radii of $5''$ and $15''$), yielding $r$-band magnitudes of 24.5 ($0.4\sigma$) and 21.8 ($1.1\sigma$), respectively. In all cases, no significant signal is detected at the center. This result indicates that the surface brightness of the dark galaxy candidates falls below the detection limit of the stacked image, or that the optical counterparts are more compact than the positional uncertainties. Deeper optical follow-up observations are needed to determine whether these candidates have optical counterparts.

\section{Conclusions}\label{sec:conclusion}
We use the Arecibo Legacy Fast ALFA (ALFALFA) untargeted {\HI} survey to construct a large sample dark galaxy candidates in the local universe. We then compare their physical properties with those of luminous galaxies in observations and with those of dark galaxy candidates in simulations. Our primary findings can be summarized as follows:
\begin{enumerate}
    \item Starting from 344 {\HI} sources without optical counterparts in \cite{2018ApJ...861...49H}, we could obtain a final sample of 142 dark galaxy candidates. Their physical properties (i.e. mass, {\HI}-to-dynamical mass ratio, local density) are similar to those from cosmological simulations.
    \item The dark galaxy candidates show different characteristics from the luminous galaxies. The dynamical and {\HI} masses of dark galaxy candidates are generally lower than those of luminous galaxies. They also have higher {\HI}-to-dynamical mass ratios than luminous galaxies.
    \item The dark galaxy candidates are primarily found in less dense regions compared to luminous galaxies. This tendency remains even after accounting for mass selection bias between the samples.
\end{enumerate}
Upcoming surveys including the Square Kilometre Array (SKA), EUCLID and LSST will enable us to validate our current sample and to identify a larger number of dark galaxy candidates. These efforts will enhance our understanding of dark galaxies and  provide deeper insights into galaxy formation and evolution models. 

\begin{acknowledgments}
We thank the referee for constructive comments that improved the paper.
We thank many members of the ALFALFA team, particularly Martha P. Haynes, for making the survey.
HSH acknowledges the support of the National Research Foundation of Korea (NRF) grant funded by the Korea government (MSIT), NRF-2021R1A2C1094577, Samsung Electronic Co., Ltd. (Project Number IO220811-01945-01), and Hyunsong Educational \& Cultural Foundation. 
The National Radio Astronomy Observatory is a facility of the National Science Foundation operated under cooperative agreement by Associated Universities, Inc.
This work was supported by the National Research Foundation of Korea (NRF) grant funded by the Korea government (MSIT) (RS-2025-00516062).
\end{acknowledgments}
\vspace{5mm}


\def\arraystretch{1.2}
\begin{table*}
\centering
\movetabledown=67mm
\begin{rotatetable*}
\renewcommand\thetable{1}
\caption{Dark galaxy candidates catalog}
\begin{tabular}{ccccccccccccccccc}
\hline
\hline
AGC ID& Name & R.A. & DEC. & $cz_{\odot}$ & $D_{HI}$ & SNR & $W_{50}$ & $W_{20}$ & $\rm \log M_{HI}$ &$\rm \log M_{ dyn}$ &Asym&$\rho_{20}/\bar{\rho}$ & LR & $d_{min}$&code&note\\
& & $(\rm deg)$ & $(\rm deg)$ & $(\rm km \, s^{-1})$ & $(\rm Mpc)$ &  & $(\rm km \, s^{-1})$ & $(\rm km \, s^{-1})$ & $\rm (\log M_{\odot})$  &$\rm (\log M_{\odot})$ & & & & $(\rm Mpc)$ & &  \\

(1)&(2)&(3)&(4)&(5)&(6)&(7)&(8)&(9)&(10)&(11)&(12)&(13)&(14)&(15)&(16)&(17)\\
\hline
\decimals
208745 & nrN3227. & 156.4970 & 20.3642 & 1219 & 20.2(2.1) & 63.6 & 46 & 68 & 8.52(0.10) & 9.15(0.41) & 0.059 & 3.312 & 0.020 & 0.042 & 1 & [L2016]\\
208746 & nrI610.. & 156.6238 & 20.3717 & 1241 & 20.5(2.3) & 13.2 & 67 & 89 & 8.05(0.11)& 9.14(0.41) & 0.027 & 2.988 & 0.050 & 0.043 & 1 & [L2016]\\
208522 & ........ & 156.8317 & 1.2536 & 6753 & 101.6(2.2) & 16.5 & 37 & 71 & 9.54(0.05)  & 9.70(0.44) & 0.043 & 0.198 & 0.028 & 0.215 & 4 & near U5667\\
208881 & ........ & 157.5758 & 31.2872 & 4467 & 65.7(2.2) & 9.6 & 28 & 39 & 8.79(0.06)  & 8.80(0.43) & 0.106 & 0.123 & 0.031 & 0.731 & 0 & \\
208428 & ........ & 158.2838 & 7.3944 & 6658 & 100.1(2.4) & 8.0 & 105 & 141 & 9.31(0.06)  & 10.18(0.47) & 0.027 & 0.072 & 0.190 & 0.512 & 5 & \\
208430 & ........ & 159.3779 & 6.9642 & 11376 & 167.6(2.3) & 8.1 & 56 & 94 & 9.58(0.06)  & 9.96(1.01) & 0.244 & 0.075 & 0.021 & 3.690 & 0 & \\
208431 & ........ & 159.8104 & 7.3522 & 3722 & 56.0(2.5) & 7.1 & 35 & 54 & 8.49(0.08)  & 8.93(0.59) & 0.074 & 0.084 & 0.046 & 1.029 & 0 & \\
205287 & LeoRing. & 161.6513 & 12.6264 & 957 & 11.1(2.4) & 37.6 & 78 & 122 & 8.01(0.19)  & 9.40(0.45) & 0.063 & 1.849 & 0.050 & 0.026 & 1 & [S2009]$\&$[L2016]\\
205289 & LeoRing. & 161.6529 & 12.4317 & 1006 & 11.1(2.2) & 56.9 & 48 & 72 & 8.12(0.18)  & 8.99(0.43) & 0.003 & 1.828 & 0.011 & 0.038 & 1 & [S2009]$\&$[L2016]\\
205290 & LeoRing. & 161.6779 & 12.7797 & 915 & 11.1(2.3) & 19.1 & 50 & 87 & 7.68(0.18)  & 8.94(0.46) & 0.111 & 1.869 & 0.023 & 0.012 & 1 & [S2009]$\&$[L2016]\\
\hline
\multicolumn{15}{c}{Table 1 is published in its entirety in the machine-readable format. A portion is shown here for guidance regarding its form and content.}
\end{tabular}
\label{tab:1}
\end{rotatetable*}
\end{table*}

\end{document}